\documentclass[useAMS,usenatbib]{mn2e}
\usepackage{graphicx}

\title[Non-axisymmetric solar dynamo]{A dynamo model for axisymmetric and
 non-axisymmetric solar magnetic fields}
\author[J. Jiang and J. X. Wang]{J. Jiang$$\thanks{E-mail:
jiangjie@ourstar.bao.ac.cn (JJ);
 wangjx@bao.ac.cn}
and J. X. Wang$$\footnotemark[1]\
\\
$$National Astronomical Observatories, Chinese Academy of
Sciences, Beijing 100012, China}
\begin{document}

\date{Accepted 0000 xxxx 00. Received 0000 xxxx 00; in original form 0000 xxxx 00}

\pagerange{\pageref{firstpage}--\pageref{lastpage}} \pubyear{0000}

\maketitle

\label{firstpage}

\begin{abstract}
Increasing observations are becoming available about a relatively
weak, but persistent, non-axisymmetric magnetic field co-existing
with the dominant axisymmetric field on the Sun. It indicates that
the non-axisymmetric magnetic field plays an important role in the
origin of solar activity. A linear non-axisymmetric
$\alpha^2-\Omega$ dynamo model is set up to discuss the
characteristics of the axisymmetric ($m=0$) and the first
non-axisymmetric ($m=1$) modes and to provide further the
theoretical bases to explain the `active longitude', `flip-flop'
and other non-axisymmetric phenomena. The model consists of a
updated solar internal differential rotation, a turbulent
diffusivity varied with depth and an $\alpha$-effect working at
the tachocline in rotating spherical systems. The difference
between the $\alpha^2-\Omega$ and the $\alpha-\Omega$ models and
the conditions to favor the non-axisymmetric modes with the
solar-like parameters are also presented.
\end{abstract}


\section{Introduction}

The distribution of magnetic field emerging on the solar surface
carries clues to the mechanism of the field generation. One
striking feature of this distribution is clustering of active
regions which is commonly called `active longitude' \citep{bai87,
ben99, det00}. Signatures of possible longitudinal inhomogeneities
have also been reported in the distributions of solar wind and
interplanetary magnetic field \cite[]{neu00}. Furthermore,
`flip-flop' phenomenon, i.e. the two persistent active longitudes
separated by $180\degr$, has also been identified on the Sun
\cite[]{ber03}. These observations indicate the involvement of
large-scale non-axisymmetric magnetic field in the formation and
evolution of the dominant axisymmetric solar activities. Hence, it
is valuable to set up the non-axisymmetric dynamo model to explain
these non-axisymmetric solar magnetic fields.

The pioneer works on the theoretical investigations of the
non-axisymmetric activities can be classed mainly as two kinds.
One is that the generation sources are non-axisymmetric, and the
non-axisymmetric magnetic field is produced accordingly. For
example, Bigazzi \& Ruzmaikin (2004) and Moss et al. (2002)
adopted the non-axisymmetric distribution of $\alpha$-effect. The
other is based on the axisymmetric sources of generation but to
excite the non-axisymmetric field. The numerical results of Chan
et al. (2004) supported this possibility.

Earlier studies \citep{sti71, iva85} concerning the linear
non-axisymmetric solar dynamo with decoupled axisymmetric and
non-axisymmetric modes have been taken. But these earlier studies
could not include the correct distribution of solar differential
rotation, which was unknown at that time. Recently, there are some
works on the non-linear non-axisymmetric dynamo models. Moss
(1999) obtained stable solutions which possessed a small
non-axisymmetric field component co-existing with a dominant
axisymmetric part with the updated solar rotation profile. Bigazzi
\& Ruzmaikin (2004) studied the generation of non-axisymmetric
fields and their coupling with the axisymmetric solar magnetic
field. Bassom et al. (2005) used an asymptotic WKBJ method to
investigate a linear $\alpha^2-\Omega$ model with the aim to
isolate the basic physical effects leading to the preferable
excitation of non-axisymmetric solar and stellar magnetic
structure.  However, are there possibilities to work out a linear
non-axisymmetric solar dynamo with the updated generation sources
to relate with the non-axisymmetric phenomena? What are the
differences, such as configuration and cycle and so on, between
the axisymmetric and non-axisymmetric modes? When will the
non-axisymmetric mode be preferred? These are the main objectives
of the paper.

With the axisymmetric sources of generation, we develop a new
high-precision non-axisymmetric code based on the spectral method
and begin with the linear non-axisymmetric mean field dynamo
equations. The axisymmetric mode $m=0$ and the first
non-axisymmetric mode $m=1$ are discussed, respectively, in Sec.5
and Sec.6. We will show the difference between the
$\alpha^2-\Omega$ and the $\alpha-\Omega$ models in Sec.3. In Sec.
4, the condition to excite the dominant axisymmetric mode and the
condition to favor the non-axisymmetric mode will be discussed.

\section{Mathematic Formulations}
\subsection{The basic equations}
The starting point of our model is the mean field dynamo equation,
governing the evolution of the large-scale magnetic field
\emph{\textbf{B}} in response to the flow field \emph{\textbf{U}},
the $\alpha$-effect and the magnetic diffusivity $\eta$:
\begin{equation}
\frac{\partial\textbf{\emph{B}}}{\partial\emph{t}}=\nabla\times[\emph{\textbf{U}}\times
\emph{\textbf{B}}+\alpha\textbf{\emph{B}}
-\eta\nabla\times\textbf{\emph{B}}].
\end{equation}
Since the turbulent diffusivity is much larger than the molecular
diffusivity, we ignore the molecular diffusivity in $\eta$. For
the flow field, only the (differential) rotation $\Omega$ is
considered for simplicity. Since the magnetic field is
divergence-free, we expand $\textbf{\emph{B}}$ in term of two
scalar functions \emph{h} and \emph{g} which represent the
poloidal and toroidal potentials, respectively, in the spherical
polar coordinates $(r,\theta,\phi)$ as \cite{cha61} and
\cite{mof78}.
\begin{equation}
\textbf{\emph{B}}=\nabla\times\nabla\times\emph{\textbf{r}}\emph{h}
(r,\theta,\phi,t)+\nabla\times\textbf{\emph{r}}\emph{g}(r,\theta,\phi,t).
\end{equation}
When $\alpha=\alpha(r,\theta)$, $\Omega=\Omega(r,\theta)$,
$\eta=\eta(r)$, substituting equation(2) in equation (1), the
governing equation reduces to:
\begin{eqnarray}
\partial_{t}L^{2}h&=&R_{\alpha}V^{g}_{\alpha}+\eta\nabla^{2}L^{2}h+R_{\Omega}V^{h}_{\Omega
N}+R_{\alpha}V^{h}_{\alpha N},\\
\partial_{t}L^{2}g&=&R_{\alpha}V^{h}_{\alpha}+R_{\Omega}V^{h}_{\Omega}+
R_{\Omega}V^{g}_{\Omega N}+R_{\alpha}V^{g}_{\alpha N} \nonumber\\
&+& \eta\nabla^{2}L^{2}g+\frac{\partial\eta}{\partial
r}\frac{\partial}{\partial
r}L^{2}g+\frac{1}{r}\frac{\partial\eta}{\partial r}L^{2}g,
\end{eqnarray}
where
\begin{displaymath}
L^{2}=-\frac{1}{\sin\theta}\frac{\partial}{\partial\theta}(\sin\theta\frac{\partial}{\partial\theta})-
\frac{1}{\sin^{2}\theta},
\nabla^2=\frac{1}{r}\frac{\partial}{\partial
r}r^2\frac{\partial}{\partial r}-\frac{1}{r^2}L^2.
\end{displaymath}
 $V_{\Omega N}$ and $V_{\alpha N}$ are the terms which have the
azimuthal component $\partial/\partial \phi$ and $V_{\Omega}$,
$V_{\alpha},V_{\Omega N}$ and $V_{\alpha N}$ can be obtained from
(see Appendix):
\begin{eqnarray}
\emph{\textbf{r}}\cdot\nabla\times(\alpha\emph{\textbf{B}})&=&V^{g}_{\alpha}+V^{h}_{\alpha
N},\\
\emph{\textbf{r}}\cdot\nabla\times(\emph{\textbf{U}}\times\emph{\textbf{B}})&=&V^{h}_{\Omega
N},\\
\emph{\textbf{r}}\cdot\nabla\times[\nabla\times(\alpha\emph{\textbf{B}})]&=&V^{h}_{\alpha}+V^{g}_{\alpha
N},\\
\emph{\textbf{r}}\cdot\nabla\times[\nabla\times(\emph{\textbf{U}}\times\emph{\textbf{B}})]
&=&V^{h}_{\Omega}+V^{g}_{\Omega N}.
\end{eqnarray}
The equations (3) and (4) have been cast in non-dimensional form
by expressing all lengths in units of solar radius $R_\odot$ and
time in units of the magnetic diffusion time $R_\odot^{2}/\eta_o$.
This has led to the appearance of two dimensionless numbers:
\begin{eqnarray}
R_\alpha&=&\frac{\alpha_o R_\odot}{\eta_o},\\
R_\Omega&=&\frac{\Omega_o R_\odot^{2}}{\eta_o},
\end{eqnarray}
where $\alpha_o$ and $\eta_o$ are reference values for the
$\alpha$-effect and the diffusivity in the convective zone (CZ),
respectively. And $\Omega_o$ is the characteristic value of the
differential rotation. The quantities $R_\alpha$ and $R_\Omega$
are dynamo numbers measuring the relative importance of inductive
versus diffusive effects. More discussion about $R_\Omega$ will be
given in Subsection 2.2.

\subsection{Internal rotation $\Omega(r,\theta)$}
Based on the helioseismic inversion \citep{sch98,cha99}, there are
two strong radial shear regions inside of the Sun. One is in the
tachocline and the other locates in sub-photospheric layer. For
the sake of simplicity on computational solutions, We neglect the
shear at the sub-surface and regard that the dynamo works in the
tachocline. The following expression for the solar interior
rotation is adopted.
\begin{eqnarray}
\Omega(r,\theta)=\Omega_c+\frac{1}{2}[1+erf(2\frac{r-r_c}{d})](\Omega_s(\theta)-\Omega_c),
\end{eqnarray}
where
$\Omega_s(\theta)=\Omega_{EQ}+a_2\cos^{2}\theta+a_4\cos^{4}\theta$
is the surface latitudinal rotation and $\theta$ is co-latitude.
The parametric values are set as $r_c=0.7R_\odot$,
$d=0.05R_\odot$, $\Omega_c/2\pi=430.0$~nHz,
$\Omega_{EQ}/2\pi=455.8$ nHz, $a_2/2\pi=-51.2$ nHz,
$a_4/2\pi=-84.0$ nHz.  Fig. 1 shows the radial distribution of
$\Omega(r,\theta)$ at different latitudes. It reveals that
$\Omega$ depends weakly on depth in bulk of CZ. But in the
tachocline, the rotation rate changes from almost uniform in the
radiative interior to depth dependent in the CZ. Within the
tachocline, rotation increases with distance from the core at low
latitudes, while it decreases at high latitudes. At intermediate
latitudes (near $35\degr$, dashed line in Fig. 1) rotation is
almost independent on the depth.

Furthermore, we base our model on the rotating spherical systems
with the rotation velocity $\Omega_c$ of the inner core. Thus the
differential rotation in the rotating frame $\Omega_c$ is
\begin{eqnarray}
\Omega'(r,\theta)&=&\frac{1}{2}[1+erf(2\frac{r-r_c}{d})](2\pi\times25.8)\times\nonumber\\
& &(1.-1.98\cos^{2}\theta-3.26\cos^{4}\theta)~ \rm{(nHz)}.
\end{eqnarray}
The differential rotation of the surface at the equator is
($2\pi\times25.8$) nHz and we regard it as the characteristic
value of the differential rotation $\Omega_o$ in Eq. (10). Hence,
the value of $R_\Omega$ is adopted as $\frac{\displaystyle
8\times10^{10}~\rm{m^2 s^{-1}}}{\displaystyle\eta_o}$, which is
only decided by the reference value of the diffusivity $\eta_o$.

\begin{figure}
  \centering
\includegraphics[width=65mm]{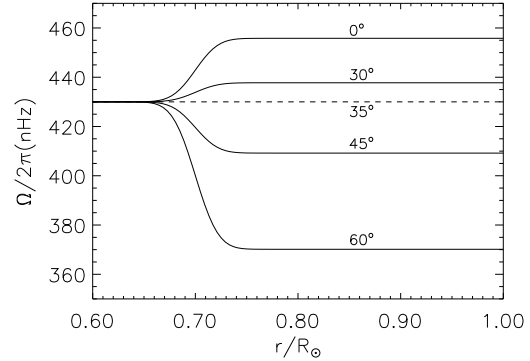}
\caption{Radial distributions of solar rotation at different
latitudes. There are strong radial shear in the high latitude of
the tachocline. At $35\degr$ latitude, the shear is very weak
(dashed line).}
\end{figure}

\begin{figure}
  \centering
\includegraphics[width=65mm]{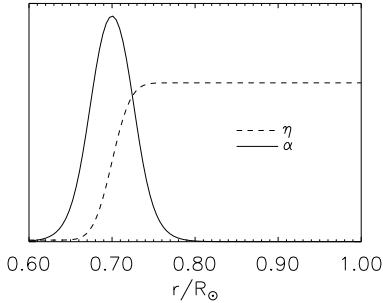}
\caption{Radial distributions of the $\alpha$-effect (solid line)
which mainly concentrate in the tachocline and the magnetic
diffusivity $\eta$ (dashed line). Values of y-axis are not
scaled.}
\end{figure}

\subsection{The diffusivity profile $\eta(r)$}
We use the analytical expression of Dikpati \& Charbonneau (1999)
for the diffusivity profile as
\begin{equation}
\eta(r)=\eta_c+\frac{\eta_o}{2}[1+erf(2\frac{r-r_c}{d})],
\end{equation}
whose distribution can be seen in Fig. 2 (dashed line). The
diffusivity $\eta_o$ in CZ, is dominated by the turbulence. In the
stably stratified core, the diffusivity $\eta_c$ is much lower
because of the much less turbulence. In what follows we take
$\eta_c/\eta_o$=0.01. The transition from high to low diffusivity
occurs near the tachocline, which is coincident with the
rotational shear layer. Here, $\eta_o$ is far less definite and is
widely known to fall in the range from $2\times
10^{10}~\rm{cm^2s^{-1}}$ to $2\times 10^{12}~\rm{cm^2s^{-1}}$.

\subsection{The $\alpha$-effect $\alpha(r,\theta)$}
The $\alpha$-effect cannot yet be determined from observations.
The dominated physical mechanisms responsible for it can be
categorized as the following three types. (1) It works at the
surface produced by the decaying of active regions \citep{bab61,
lei69}. (2) It is directly related to turbulent convective motions
\cite[]{par55}. It exists throughout the whole CZ and changes sign
near the bottom of the CZ \citep{kri98, kuz03}. (3) It works at
the tachocline induced by the hydrodynamical shear instabilities
\cite[]{dik01} or MHD instabilities \cite[]{the00}. It is possible
that all of them work simultaneously inside of the Sun. Here, we
only consider the $\alpha$-effect concentrating in the tachocline
with the following expression
\begin{eqnarray}
\alpha(r,\theta)&=&\alpha_o~\frac{1}{2}[1+erf(2\frac{r-r_1}{d})]\nonumber\\
& &~~~~\frac{1}{2}[1-erf(2\frac{r-r_2}{d})]\cos\theta,
\end{eqnarray}
where $r_1=0.675R_\odot$, $r_2=0.725R_\odot$. The solid line in
Fig. 2 shows the variation of $\alpha(r,\theta)$ with $r$, which
is mainly concentrated in the tachocline. The common angular
dependence $\cos\theta$ is adopted, the simplest guaranteeing
antisymmetry across the equator. Moreover, we do not consider the
$\alpha$-quenching since only the linear solutions are sought.

\subsection{The numerical scheme}
Since the governing equations (3) and (4) are two coupled, linear
homogeneous equations in $h$ and $g$, with the given boundary
conditions, we can look for their eigensolutions with the form
 \begin{equation}
[\emph{h}(\emph{r},\theta,\phi,t),\emph{g}(\emph{r},\theta,\phi,t)]=
[\emph{h}(\emph{r},\theta,\phi),\emph{g}(\emph{r},\theta,\phi)]\mathrm{e}^{st},
\end{equation}
where \emph{s} is the eigenvalue and can be written as
$\emph{s}=\sigma+\mathrm{\emph{i}}\omega$. Only the solution that
neither grows nor decays ($\sigma\simeq 0$), i.e. the onset of
dynamo actions is considered. The corresponding $R_\alpha$ is the
critical $R_\alpha$. The solution with the lowest dynamo number is
the easiest to excite and is the most stable one. In what
immediately follows, we solve the dynamo equations numerically
using spectral (Chebyshev-$\tau$) method (see Jiang \& Wang 2006
for detail).

Different azimuthal modes $m$ are decoupled in linear theory. For
given $m$, we expand $h$ and $g$ at the onset of the dynamo action
in terms of Chebyshev polynomial $T_{n}(r)$ and surface harmonics
$P_l^m \mathrm{e}^{\mathrm{i} m\phi}$ in the meridional circular
sector $r\in[0.6,1.0]$, $\theta\in[0.0,\pi]$ as follows:
\begin{eqnarray}
 h=\sum_{n=0}^{N}\sum_{l=m}^{L}c_{n,l}^h
 T_n(a r-b)P_l^m(\cos\theta)\mathrm{e}^{\mathrm{i}m\phi},\\
 g=\sum_{n=0}^{N}\sum_{l=m}^{L}c_{n,l}^g
 T_n(a r-b)P_l^m(\cos\theta)\mathrm{e}^{\mathrm{i}m\phi},
\end{eqnarray}
where $a r-b\in[-1,+1]$. Here, $N$ and $L$ are the truncations
needed to get convergence. It varies with different dynamo number
and different $\Omega$, $\eta$, $\alpha$ profiles. $c_{n,l}^h$ and
$c_{n,l}^g$ are eigenvectors.

At two interface $r=r_o=1.0$ and $r=r_i=0.6$, both magnetic field
and the tangential electric field must be continuous. The exterior
$r>1.0$ is a vacuum and eigensolutions are matched to a potential
field. The radiative core is assumed to behave as a perfect
conductor. We may obtain (see Schubert \& Zhang 2001 for detail):
\begin{eqnarray}
\rm{at} ~~r=r_i=0.6,~~\sum_{n}\sum_{l}c_{n,l}^h(-1)^n P_l^m(\cos\theta)&=&0,\\
\sum_{n}\sum_{l}c_{n,l}^h[a(-1)^{n+1}n^2r_i+(-1)^n]P_l^m(\cos\theta)&=&0,\\
\rm{at} ~~r=r_o=1.0,~~\sum_{n}\sum_{l}c_{n,l}^gP_l^m(\cos\theta)&=&0,\\
~~\sum_{n}\sum_{l}c_{n,l}^h [a n^2r_o+(1+l)]P_l^m(\cos\theta)&=&0.
\end{eqnarray}

As pointed out by Ivanova \& Ruzmaikin (1985), the system of Eqs.
(3) and (4) may be decomposed into two subsystems, i.e. odd or
even parity with respect to the equatorial plane. We denote them
by $A$ and $S$. With the parameters adopted in our model, both odd
and even parity solutions have nearly the same dynamo number for
given mode $m$. Therefore, we cannot decide which kind of
symmetric solution is excited easier. However, according to the
observations and theoretical computations \citep{mos04,flu04}, the
dipolar mode ($A0$) for axisymmetric field and the perpendicular
dipolar mode ($S1$) for non-axisymmetric field are definitely
identified on the Sun although there are some possibilities for
some other non-axisymmetric modes exist on the Sun \citep{det00,
son05}. Hence we will only choose the two modes $A0$ and $S1$ as
the representatives of axisymmetric and non-axisymmetric modes,
respectively and investigate $A0$ and $S1$ in detail below. We
firstly discuss the difference between the $\alpha^{2}-\Omega$ and
$\alpha-\Omega$ models and then give the condition for the Sun to
excite the preferred non-axisymmetric mode.

\begin{figure}
  \centering
\includegraphics[width=65mm]{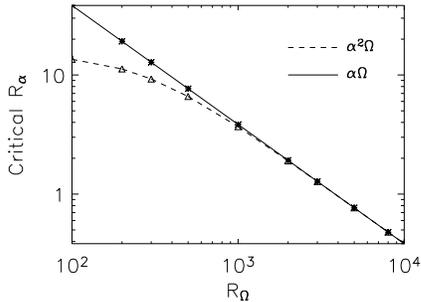}
\caption{A critical $R_\alpha$-$R_\Omega$ plot for the
$\alpha-\Omega$ (solid line) and the $\alpha^2-\Omega$ (dashed line)
models for the axisymmetric mode $A0$. When $R_\Omega\leq
3\times10^3$, the $\alpha-\Omega$ model has larger critical
$R_\alpha$ than the $\alpha^2-\Omega$ one. When
$R_\Omega>3\times10^3$, the two models have the nearly same critical
$R_\alpha$. For the $\alpha-\Omega$ model, $R_\alpha R_\Omega$ is
about 3840.}
\end{figure}

\begin{figure}
  \centering
\includegraphics[width=65mm]{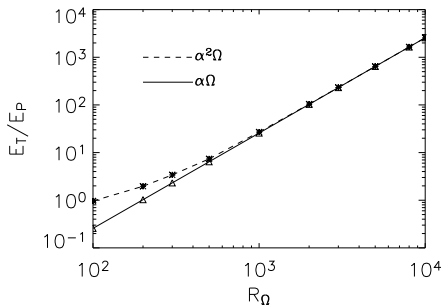}
\caption{The ratio of the magnetic energy between the toroidal and
poloidal components vs. $R_\Omega$ for the $\alpha-\Omega$ (solid
line) and the $\alpha^2-\Omega$ (dashed line) models for the mode
$A0$.}
\end{figure}

\section{The $\alpha^2-\Omega$ Dynamo Model Versus the $\alpha-\Omega$ One}
For the mean field dynamo theory, poloidal field is created from
toroidal field by the $\alpha$-effect and toroidal field from
poloidal field by two ways, i.e. the differential rotation
($\Omega$-effect) and $\alpha$-effect. The model including all the
ingredients is called the $\alpha^2-\Omega$ model. When
$R_\Omega\gg R_\alpha^2$, the $\alpha$-effect as the toroidal
source can be ignored. The $\alpha-\Omega$ model is always adopted
\cite[]{zel83}. Is the simple $\alpha-\Omega$ model fine for the
Sun? What quantitative conditions does it need to satisfy? What
are the definite differences between the two models with
solar-like parameters?

We first discuss them based on the axisymmetric mode $A0$ and
enlarge the range of $\eta_o$ from $8\times10^{10} ~\rm{cm^2s^{-1}}$
to $8\times10^{12} ~\rm{cm^2s^{-1}}$. Thus we obtain $R_\Omega$
ranging from $10^2$ to $10^4$. For the $\alpha-\Omega$ model, the
condition for the generation of undamped magnetic field is only
determined by $D=R_\alpha R_\Omega$ \cite[]{iva85}. Accordingly we
obtain the straight (solid) line in Fig. 3 with logarithmic abscissa
and $D=R_\alpha R_\Omega$ is about 3840. But for the
$\alpha^2-\Omega$ model, it is more complicated with the generation
of toroidal field by the $\alpha$-effect (the dashed line in Fig.3).
Comparing the two lines of Fig. 3, we can see that for the
$\alpha^2-\Omega$ model, dynamo action is increased contrasting with
the $\alpha-\Omega$ model by the reduction of the critical
$R_\alpha$ when $R_\Omega$ is small ($<3\times 10^3$). With the
increasing of $R_\Omega$, the difference for the corresponding
critical $R_\alpha$ between the two models decreases. When
$R_\Omega=3\times 10^3$, the agreement between the two models
reaches the level of 0.3\%.

Since the absolute scale for the strength of magnetic field is
undermined by linear eigenvalues calculations, we define the ratio
of the magnetic energy between the toroidal and poloidal
components as \cite[]{cha01}:
\begin{equation}
 \Theta=\frac{\int B_T^2~dV}{\int B_P^2~dV}~,
\end{equation}
where $\emph{\textbf{B}}_T=-\partial g/\partial \theta~
\textbf{\^{e}}_\phi$ and
$\emph{\textbf{B}}_P=\displaystyle\frac{L^2h}{r}~\textbf{\^{e}}_r+(\frac{1}{r}\frac{\partial
h}{\partial\theta}+\frac{\partial^2 h}{\partial
r\partial\theta})~\textbf{\^{e}}_\theta$ for the axisymmetric
model. Fig. 4 gives the energy ratios between the toroidal and
poloidal fields at the onset state for the two models with
different $R_\Omega$. When $R_\Omega$ is less than $3\times 10^3$,
the $\alpha-\Omega$ model produces smaller $E_T/E_P$ than the
$\alpha^2-\Omega$ model. The larger $R_\Omega$ is, the less
difference the two models have. The agreement between the two
models reaches 0.46\% with $R_\Omega=3\times 10^3$. When
$R_\Omega>3\times 10^3$, the energy ratios are closely in
conformity with each other. Thus we can replace the
$\alpha^2-\Omega$ model by the $\alpha-\Omega$ model and the
corresponding turbulent diffusivity $\eta_o$ should be less than
$2.67\times 10^{11}~\rm{cm^2s^{-1}}$.

Let us see the non-axisymmetric mode $S1$ simply now. Table 1
gives the difference for the two models with different $R_\Omega$.
When $R_\Omega>4\times 10^3$ ($\eta_o < 2\times 10^{11}
\rm{cm^2s^{-1}}$), the difference between the two kinds of models
is 0.73\% and the two models can be regarded as the same. In a
word, if the non-axisymmetric mode $m=1$ is considered in the
model, it is necessary for $R_\Omega$ to be larger than 4000,
namely $\eta_o$ is less than $2\times 10^{11} \rm{cm^2s^{-1}}$  so
that the $\alpha-\Omega$ model is at the limit of
$\alpha^2-\Omega$ model. It is fully reasonable to adopt the
$\alpha-\Omega$ model to replace the $\alpha^2-\Omega$ one.

\begin{figure}
  \centering
\includegraphics[width=65mm]{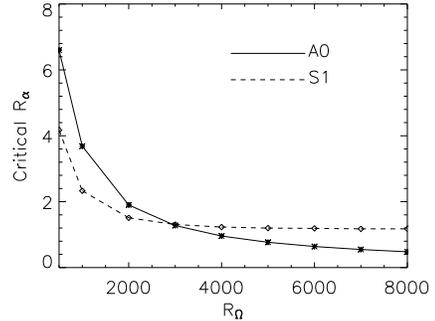}
\caption{A critical $R_\alpha$-$R_\Omega$ plot for the
$\alpha^2-\Omega$ model with the modes $A0$ (solid line) and $S1$
(dashed line). When $R_\Omega\leq 3\times 10^3$, the mode $S1$ has
smaller $R_\alpha$. It is contrary for $R_\Omega> 3\times 10^3$ that
the mode $A0$ owns smaller $R_\alpha$.}
\end{figure}

\begin{table*}
  \caption[]{The critical $R_\alpha$ for the $\alpha^2-\Omega$
  and the $\alpha-\Omega$ models based on the first non-axisymmetric mode $S1$.
  $d=(a2-a1)/a2$. $a1$ is the critical $R_\alpha$ for the $\alpha-\Omega$ model and
  $a2$ for the $\alpha^2-\Omega$ model. }
  \begin{center}\begin{tabular}{cccccccccc}
  \hline\
$R_\Omega$ & 500 & 1000 & 2000 & 3000 & 4000 & 5000 & 6000 & 7000 & 8000 \\
  \hline
$a1$  & 4.92   & 2.45 & 1.545 & 1.320 & 1.239 & 1.203 & 1.190 & 1.180 & 1.179\\
$a2$  & 4.17   & 2.33 & 1.510 & 1.306 & 1.230 & 1.198 & 1.186 & 1.177 & 1.176    \\
$d$   & 18.0   & 5.15 & 2.32  & 1.07  & 0.73  & 0.42  & 0.34  &
0.25 & 0.25
\\
\hline
  \end{tabular}\end{center}
\end{table*}

\section{Axisymmetric Versus Non-Axisymmetric Mode}
It is commonly regarded that the strong differential rotation
works in favor of the axisymmetric mode. Without it, all single
main-sequence stars with outer CZs have the non-axisymmetric field
configurations \cite[]{rue94}. The non-uniform rotation produces
the observed dominant oscillatory dipolar field on the Sun.
However, is the differential rotation expressed by Eq. (12) is
enough to produce the dominant axisymmetric field? What condition
does it need to satisfy to make the non-axisymmetric mode
preferred?

With the given rotation profile (12) of the Sun, we range $R_\Omega$
from 500 to 8000. Accordingly, $\eta_o$ changes from $1.0\times
10^{11}~\rm{cm^2s^{-1}}$ to $1.6\times 10^{12}~\rm{cm^2s^{-1}}$.
Fig. 5 displays the critical $R_\alpha$ with different $R_\Omega$
for the $\alpha^2-\Omega$ model. The axisymmetric mode $A0$ (solid
line) has lower critical $R_\alpha$ when $R_\Omega > 3000$ and will
be preferred to excite. It is contrary for $R_\Omega < 3000$ that
the non-axisymmetric mode $S1$ will have lower critical $R_\alpha$
and will be the preferred mode. The smaller $R_\Omega$ is, the
easier $S1$ is excited. According to Sec. 3, it is not at the
$\alpha-\Omega$ limit ($R_\Omega < 3000$) when $S1$ is the preferred
mode. In other words, it is impossible to favor the non-axisymmetric
mode at the $\alpha-\Omega$ limit. To get the preferred
non-axisymmetric modes, the contribution of $\alpha$-effect to
generation of the toroidal field cannot completely vanish
\cite[]{rae86a}. The farther it deviates from the limit, the more
important roles the $\alpha$-effect plays to produce the toroidal
field and the easier the non-axisymmetric mode to excite. This is
also consistent with the analytical results of Bassom et al. (2005).
Furthermore, they gave the reason that the wind-up of
non-axisymmetric structures can be compensated by phase mixing
inherent to the $\alpha^2-\Omega$ dynamo.

Moreover, rather than the differential rotation, $R_\Omega$ is the
decisive parameter to decide which kind of mode is preferred. We may
also say that the turbulent diffusivity $\eta_o$ is the key
parameter since the differential rotation has been basically
determined by the observation \cite[]{jia07}. For the
$\alpha^2-\Omega$ model, when $R_\Omega < 3000$, i.e. $\eta_o >
2.67\times 10^{11}~\rm{cm^2s^{-1}}$ the non-axisymmetric mode will
be preferred.

In the coming two sections, we will take the $\alpha^2-\Omega$
model with $\eta_o$=$1.6\times 10^{11}~\rm{cm^2s^{-1}}$ and
$R_\Omega$=5000 so that the axisymmetric mode will be preferred.
This is the real picture of the Sun.

\begin{figure}
  \centering
\includegraphics[width=85mm]{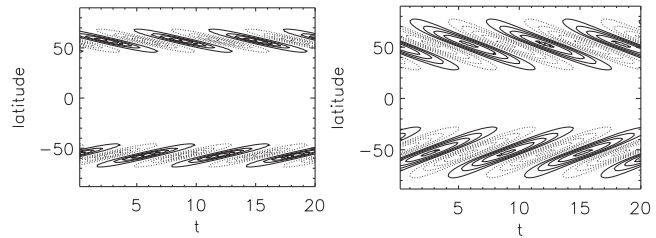}
\caption{Butterfly diagram of the toroidal field $B_\phi$ for the
mode $A0$ at the depth $r=0.7R_\odot$ (left) and $B_r$  at the
surface $r=R_\odot$ (right). Solid (dashed) contours correspond to
positive (negative) magnetic field. The diffusion time
$R_\odot/\eta_o$ is taken as the time unit.}
\end{figure}

\begin{figure}
  \centering
\includegraphics[width=80mm]{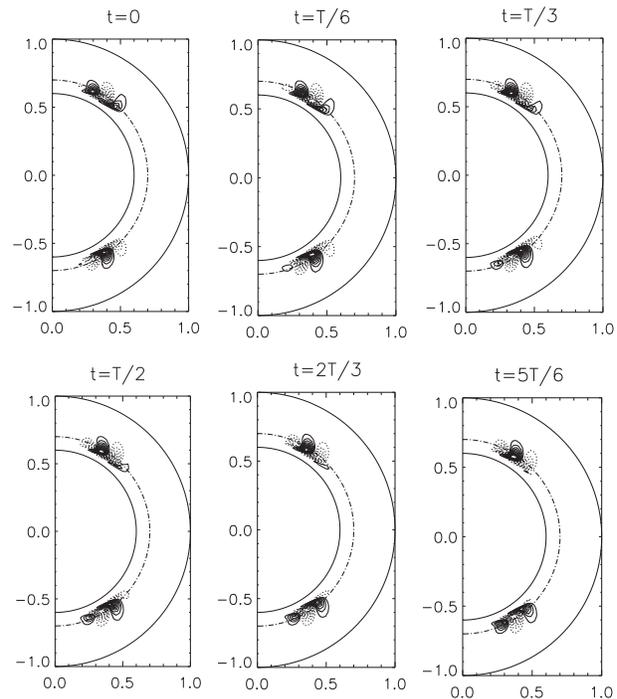}
\caption{Evolution of the toroidal magnetic field in a meridional
plane $\phi=0$ separated by 1/6th of the dynamo period $T$ for the
axisymmetric mode $A0$. Solid (dashed) contours correspond to
positive (negative) toroidal field. The dot-dashed lines locate at
$0.7R_\odot$. The fields concentrate in the tachocline and the
high latitudes where the radial shear is strong.}
\end{figure}

\section{The Axisymmetric Mode}
Left part of Table 2 is the truncation levels in the calculation of
axisymmetric mode $A0$ and the corresponding critical $R_\alpha$ and
frequency $\omega$. $N$ is the radial harmonics expanded in terms of
Chebyshev function and $L$ is the harmonics in Legendre function
(see Eqs.(16)(17)). When $N=38$ and $L=50$, it gets convergence. The
critical $R_\alpha$ is 0.765 and frequency $\omega$ is $\pm 410.37$.
With the dimensionless time $t=R_\odot^2/\eta_o$, we obtain the
period is about 15 year, which is a bit shorter than the 22-years
solar magnetic cycle. In fact, the meridional circulation omitted in
the paper plays an important role in determining the true cycle
\cite[]{dik99}. The symbol `$\pm$' corresponds to the dynamo wave
propagating equatorward or poleward.

Fig. 6 (left) displays the toroidal field at $r=0.7R_\odot$ with
$\omega=+410.37$. The magnetic field concentrates in the region
from latitude $45\degr$ to $70\degr$ with strong radial shear
$\partial \Omega/\partial r <0$ there. It is well-known that,
without meridional circulation, the propagation direction of the
dynamo wave is decided by the Parker-Yoshimura sign rule and it
should be equatorward since the product of the $\alpha$-effect and
radial gradient of differential rotation is negative
\citep{par55,yos75}. The solution with $\omega=-410.37$
corresponds to the dynamo wave propagating poleward. Although some
high-frequency dynamo waves were identified to propagate poleward
on the Sun \citep{mak89,obr06}, the poleward solution obtained in
our method cannot be used to explain these observations. It is
beyond the limitation of our method. Hence, the poleward solution
is meaningless and should be neglected. Fig. 6 (right) is a
time-latitude diagram of the radial field at the surface. The
phase shift between the two components is near $\pi/2$ which is
consistent with the observations \cite[]{she91}.

Fig. 7 is the evolution of the toroidal field in a meridional
plane $\phi=0$ at an interval of 1/6th of solar cycle period. The
dot-dashed line is $0.7R_\odot$. The magnetic field is among the
region where the radial shear of differential rotation is strong,
i.e. the high latitude of the tachocline.  This follows the
general rule that differential rotation tends to destroy any
deviation from axisymmetry and toroidal field favors to be
produced in the strong radial shear region \citep{mof78, big04}.

It seems that the location of the toroidal field produced in the
model is higher than that of the observation. Furthermore, when
the toroidal flux rope rises through the CZ to emerge, it will
have the poleward deflection further due to the effect of Coriolis
force \cite[]{cal95}. But in the model, we omit an important
ingredient, i.e. the meridional circulation. If a meridional
circulation is considered, the strong field produced within the
tachocline at high latitude will be carried to the low latitudes.
The toroidal flux entering the CZ will become buoyantly unstable
and emerge to form the active regions at the low latitudes
\cite[]{nan02}.

\begin{table*}
  \caption[]{Truncation levels and the corresponding critical $R_\alpha$
  and frequency $\omega$ for the axisymmetric mode $A0$ (left) and the first
  non-axisymmetric mode $S1$ (right). $N$ and $L$ are the harmonics
  in Chebyshev function and Legendre function respectively.}
    \begin{minipage}[t]{80.mm}
  \centering
  \begin{tabular}{ccccc}
  \hline
$N$ &  $L$   & $R_\alpha$ & $\omega$ \\
  \hline
34  & 46 & 0.762 & $\pm 408.97$\\
36  & 46 & 0.767 & $\pm 412.06$\\
38  & 46 & 0.764 & $\pm 409.90$\\
38  & 48 & 0.765 & $\pm 410.39$\\
38  & 50 & 0.765 & $\pm 410.37$\\
\hline
  \end{tabular}
\end{minipage}
  \begin{minipage}[t]{80.mm}
  \centering
  \begin{tabular}{ccccc}
  \hline
$N$ &  $L$   & $R_\alpha$ & $\omega$ \\
  \hline
24  & 66 & 1.201 & $-0.794$\\
26  & 66 & 1.198 & $-0.919$\\
28  & 66 & 1.198 & $-0.967$\\
28  & 68 & 1.203 & $-0.946$\\
28  & 70 & 1.203 & $-1.022$\\
\hline
  \end{tabular}
\end{minipage}
\end{table*}

\section{The Non-axisymmetric Mode}
Right part of Table 2 is the truncation levels of the mode $S1$.
When $N=28$ and $L=70$, it gets convergence, which is slower than
the calculation for the axisymmetric one. The critical $R_\alpha$ is
1.203, much larger than that for the axisymmetric mode. Therefore
with taking $R_\Omega=5000$ and $\eta_o=1.6\times 10^{11}\rm {cm^2
s^{-1}}$, the axisymmetric mode will be preferred.  Fig. 8 is the
near surface distribution of the radial magnetic field for the mode
$S1$ at the fixed time. The toroidal magnetic field of the mode
$m=1$ superimposed on the axisymmetric toroidal field produces a
localized maximum (`hump'). The non-axisymmetric enhancement of the
underlying magnetic field causes the clustering of sunspots to form
`active longitudes' \cite[]{ruz01} and `Flip-flop' which behaves as
a special phenomenon of `active longitudes'.

The frequency $\omega$ is -1.022. Hence its period is nearly 400
times longer than that of the mode $A0$. Thus the non-axisymmetric
mode $S1$ appears to be rather steady or weakly oscillating
comparing to the axisymmetric mode $A0$ \cite[]{ber04}. The time
variations of the mode $A0$ are periodical. By changing the sign of
the mode $A0$, the predominant longitude jumps by about $180\degr$,
which is just the flip-flop phenomenon (see the details of Sec. 3 of
Fluri \& Berdyugina (2004)). But only based on these two modes, the
full flip-flop cycle has the same length as the $A0$ cycle rather
than the value which is a 3-4 times shorter (about 3.7 years for the
Sun) than the main activity cycle \cite[]{ber04}. There should have
more complicated field configuration working on the Sun.

Both the non-axisymmetric and axisymmetric magnetic fields are
generated by the same axisymmetric sources. They evolve
independently on each other in linear theory. In fact, some
nonlinearities, such as non-axisymmetric $\alpha$-effect
\cite[]{big04}, MHD instability \cite[]{dik05}, magnetic buoyancy
\cite[]{cha04} and the $\alpha$-quenching \citep{zha03, mos05}
induce the different modes coupled together and produce the
flip-flop cycle \cite[]{mos04}.

Left of Fig. 9 shows the contours of toroidal magnetic field in a
meridional plane. All the fields concentrate in the lower part of
the tachocline where the diffusivity is less than that of other
regions (see Fig. 2). Right of Fig. 9 displays the butterfly diagram
for the mode $S1$ at the depth $r=0.75R_\odot$. The field is mainly
concentrated around the $35\degr$ latitude, where the radial shear
is weak (see Fig. 1). This is consistent with the work of Bigazzi \&
Ruzmaikin (2004) that the non-axisymmetric field survives only in
the weak differential rotation and low diffusivity region.

\begin{figure}
  \centering
\includegraphics[width=65mm]{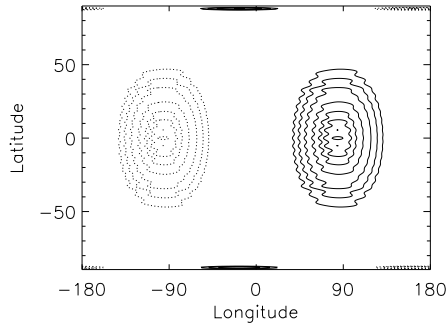}
\caption{Near surface distribution of the radial component for the
mode $S1$ at a fixed time. Solid (dashed) contours correspond to
positive (negative) magnetic field.}
\end{figure}

\begin{figure}
  \centering
\includegraphics[width=80mm]{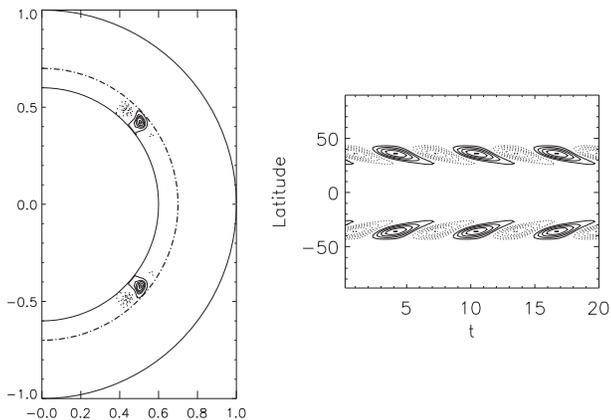}
\caption{Contours of the toroidal field $B_\phi$ in a meridional
plane (left) and Butterfly diagram of the toroidal field at the
depth $r=0.75R_\odot$ (right) for the mode $S1$. The field
concentrates in the low diffusivity (below the dot-dashed line
$0.7R_\odot$ in the left diagram) and weak differential rotation
(about the latitude $35\degr$ in the right diagram) region. Solid
(dashed) contours correspond to positive (negative) magnetic
field.}
\end{figure}

\section{Discussion and Conclusions}

In this paper we have investigated the properties of the
axisymmetric and the non-axisymmetric modes with a linear
$\alpha^2-\Omega$ model in a rotating frame trying to understand
the active longitudes, flip-flops and other non-axisymmetric
phenomena. The model consists of updated differential rotation,
turbulent diffusivity varied with depth and the $\alpha$-effect
working in the tachocline. The definite differences between the
$\alpha-\Omega$ and $\alpha^2-\Omega$ models and the conditions to
excite the non-axisymmetric modes with the solar-like parameters
are presented. The Chebyshev-$\tau$ method is used to numerically
solve the problem with a high precision. We have tested this code
with the analytical solutions of a simple $\alpha^2$ model and
compared with the results of Stix (1976). They matched well with
each other.

Although the conditions to favor the non-axisymmetric modes have
been investigated by some works \citep{rae86a,rae86b,bas05}, we
are the first to apply the updated solar parameters with numerical
method to obtain these. We also point out the role of the
turbulent diffusivity $\eta_o$ during the process. It is poorly
known and cannot be directly deduced from observations. When it is
lower than $2.0\times10^{11}~\rm{cm^2s^{-1}}$ and $R_\Omega$ is
larger than 4000 accordingly, the $\alpha-\Omega$ model is at the
limit of the $\alpha^2-\Omega$ model. Furthermore, based on the
$\alpha^2-\Omega$ model, when $R_\alpha$ is lower than 3000, the
non-axisymmetric mode $m=1$ will be more and more easily excited
than the axisymmetric mode $m=0$ when increasing the value of
diffusivity $\eta_o$ and decreasing the corresponding $R_\Omega$.

The non-axisymmetric mode $S1$ has much longer period than the
axisymmetric one $A0$. The co-existence of oscillating
 mode $A0$ and the nearly steady $S1$ results in the flip-flop phenomenon.
Because the differential rotation affects axisymmetric and
non-axisymmetric magnetic fields in the different ways, the two
kinds of fields prefer to occur in different regions. The
axisymmetric magnetic field is mainly concentrated near the high
latitude (about $55\degr$) around $0.7R_\odot$, where the radial
shear of differential rotation is strong. However, the
non-axisymmetric field occurs near the intermediate $35\degr$
latitude in the bottom of the tachocline, where the differential
rotation is weak and the diffusivity is low.

Usoskin et al. (2005) gave that the ratio between the
non-axisymmetric strength and the axisymmetric one is roughly 1:10
by analyzing sunspot group data for the past 120 years. Based on
the non-linear models, Moss (1999, 2004) presented the energy
ratios between the two kinds of modes although the values cannot
match well with the observation. Since our model is linear and
different modes are decoupled, it is the limitation to provide the
energy ratios between the two modes.

In this work, we regard the strong radial shear only exists in the
tachocline and omit the sub-surface shear and other details of the
distribution of differential rotation. Brandenburg (2005) argued
for the alternative ideas concerning dynamo operating in the bulk
of CZ, or perhaps even in the sub-surface shear layer. Moreover,
we also only take the $\alpha$-effect working in the tachocline
and ignore the other two possible mechanisms. The two generation
sources are still the hot topics on debate. Since we do not aim to
give the detailed description of the Sun and just put emphasis on
the basic characters of the axisymmetric and non-axisymmetric
modes, it is feasible for us to take the two simple generation
sources and set up the thin-layer dynamo model. Of course, more
rich and realistic models can open new option for the
understanding of solar magnetic field.

Furthermore, since we tried to expatiate on our objectives with
the simple generation sources, the meridional circulation is not
considered in the work. It plays an important role in the
axisymmetric mode \citep{nan02,gue04}. It carries the strong
axisymmetric toroidal field produced at the high latitudes to the
low ones and produces the active regions there with the magnetic
buoyancy. But it has no much influence on the non-axisymmetric
field according to the work of Bigazzi \& Ruzmaikin (2004).

In the forthcoming studies we will include the nonlinearities and
the meridional circulation to investigate the influence on the
coupling of the different modes and the role of meridional
circulation in the non-axisymmetric dynamo.

\section*{Acknowledgments}

We thank the anonymous referee for valuable comments which have
helped in improving the paper. We extend our thanks to A. R.
Choudhuri and K. M. Kuzanyan for reviewing the manuscript and giving
many helpful comments. JJ would like to thank X. H. Liao and K. K.
Zhang for kind supervision on the work. This work has been supported
by National Natural Science Foundation of China (10573025, 10603008)
and by the National Basic Research Program (G2006CB806303).

\appendix

\section{The full form of the governing equations}
The full forms of the governing equations about the toroidal field
$g$ and poloidal field $h$ in (3) and (4) are as follows:
\begin{eqnarray}
\frac{\partial L^{2}h}{\partial t}&=&R_{\alpha} [\alpha
L^{2}g-\frac{\partial \alpha}{\partial \theta}\frac{\partial
g}{\partial \theta}]+\eta\nabla^{2}L^{2}h \nonumber\\
&-& R_{\Omega}[\Omega\frac{\partial L^{2}h}{\partial\phi}]+
 R_{\alpha}\frac{1}{\sin\theta}\frac{\partial \alpha}{\partial\theta}
 [\frac{1}{r}\frac{\partial h}{\partial \phi}+\frac{\partial^{2}h}{\partial
 r\partial\phi}],
\\
\frac{\partial L^{2}g}{\partial t}&=&R_{\alpha}
[-\alpha\nabla^{2}L^{2}h+\frac{\partial
\alpha}{\partial\theta}\frac{\partial\nabla^{2}h}{\partial\theta}+
\frac{\partial^{2}\alpha}{\partial r \partial\theta}
\frac{\partial^{2}h}{\partial r \partial\theta}h \nonumber\\
&-& \frac{\partial \alpha}{\partial r}\frac{\partial
L^{2}h}{\partial r}-\frac{1}{r}\frac{\partial \alpha}{\partial r
}L^{2}h-\frac{1}{r^{2}\sin\theta}\frac{\partial}{\partial\theta}
(\sin\theta\frac{\partial\alpha}{\partial\theta})L^{2}h
\nonumber\\
&-&
\frac{1}{r^{2}}\frac{\partial\alpha}{\partial\theta}\frac{\partial
L^{2}h}{\partial\theta}+\frac{1}{r}\frac{\partial^{2} h}{\partial
r \partial\theta}]\nonumber\\
&+&
R_{\Omega}[\frac{1}{r\sin\theta}\frac{\partial}{\partial\theta}
[\sin\theta\frac{\partial\Omega}{\partial\theta}\frac{\partial}{\partial
r}r(\sin\theta\frac{\partial h}{\partial \theta})]\nonumber\\
&-&
\frac{1}{\sin\theta}\frac{\partial}{\partial\theta}[\sin\theta\frac{\partial\Omega}{\partial
r}\frac{\partial}{\partial\theta}(\sin\theta\frac{\partial
h}{\partial\theta})]
\nonumber\\
&-& \frac{1}{\sin\theta}\frac{\partial\Omega}{\partial
r}\frac{\partial^{2}}{\partial\phi^{2}}\frac{\partial
h}{\partial\theta}-\frac{1}{r\sin\theta}\frac{\partial^{2}}{\partial\phi^{2}}\frac{\partial}{\partial
r}(r\frac{\partial\Omega}{\partial\theta}h)]
\nonumber\\
&+& \eta\nabla^{2}L^{2}g+\frac{\partial\eta}{\partial
r}\frac{\partial}{\partial
r}L^{2}g+\frac{1}{r}\frac{\partial\eta}{\partial r}L^{2}g
\nonumber\\
&+&R_{\alpha}[\frac{1}{\sin\theta}\frac{\partial
g}{\partial\phi}(\frac{\partial^{2}\alpha}{\partial
r\partial\theta}+\frac{1}{r}\frac{\partial\alpha}{\partial\theta}+\frac{\partial
g}{\partial r})]
\nonumber\\
&-& R_{\Omega}L^{2}(\Omega\frac{\partial g}{\partial\phi}).
\end{eqnarray}
where $\alpha$, $\Omega$ and $\eta$ are the expressions after
non-dimensionalization.

\bsp

\label{lastpage}

\end{document}